\documentclass{elsartL}
\usepackage{graphicx}
\usepackage{amssymb}
\usepackage{amsmath}
\usepackage{mathrsfs}
\begin{document}

\begin{frontmatter}
\title{Systematic effects in the low-energy behavior of the current SAID solution for the pion-nucleon system}
\author[EM]{E.~Matsinos{$^*$}},
\author[GR]{G.~Rasche},
\address[EM]{Institute of Mechatronic Systems, Zurich University of Applied Sciences, Technikumstrasse 5, CH-8401 Winterthur, Switzerland}
\address[GR]{Physik-Institut der Universit\"at Z\"urich, Winterthurerstrasse 190, CH-8057 Z\"urich, Switzerland}

\begin{abstract}
We investigate the description of the pion-nucleon experimental data at low energy (i.e., below pion laboratory kinetic energy of $100$ MeV) on the basis of the current SAID solution (WI08). We demonstrate that, in a 
self-consistent analysis scheme, the scale factors of the fits based on the Arndt-Roper formula come out independent of the beam energy and `cluster' around the expectation value of $1$. We report systematic effects in the 
low-energy behavior of the WI08 solution. These effects indicate that at least one of the three following assumptions, underlying the WI08 analysis, does \emph{not} hold, namely that: a) the bulk of the data is reliable, b) 
the electromagnetic effects are correctly accounted for, and c) the isospin invariance in the strong interactions of the $\pi N$ system is valid.
%
%
\end{abstract}
{$^*$}{Corresponding author. E-mail: evangelos[DOT]matsinos[AT]sunrise[DOT]ch}
\end{frontmatter}

\section{\label{sec:Introduction}Introduction}

The SAID results for the pion-nucleon ($\pi N$) system \cite{abws} are widely used as input in numerous works, not only in those studying aspects of the $\pi N$ interaction. The results represent an optimization of the 
description of an extensive database (DB) of measurements, ranging from the $\pi N$ threshold to (pion laboratory kinetic energy $T$ of) a few GeV. The experimental data are analyzed via dispersion relations. The SAID 
phase-shift solution is regularly updated, conveniently appearing online, which facilitates the fast dissemination of any new results. Furthermore, new measurements are frequently communicated to the developers of this platform 
prior to their formal publication, ensuring that the site remain at the state-of-the-art level on a number of hadronic processes. Regarding the $\pi N$ interaction, the current SAID solution is known as WI08. This paper 
examines one aspect of this solution, namely its low-energy behavior. As such, it is expected to be of interest to those who extract important hadronic quantities from the low-energy SAID phase shifts or compare their 
experimental results to the predictions obtained on the basis of these phase shifts.

Regarding the SAID phase-shift solutions for $T \leq 100$ MeV, our concerns have already been expressed on several occasions. To start with, one has the impression that new $\pi N$ measurements enter the SAID DB with little 
regard for their self-consistency and/or compatibility with the measurements which are already present in the DB. Had the SAID group implemented robust statistics in their analysis, such a strategy would be less problematic; 
however, their results are obtained with a `standard' $\chi^2$ function (i.e., with the Arndt-Roper formula \cite{ar}), and are thus expected to be sensitive to the presence of outliers in the DB (in particular, of 
\emph{one-sided} outliers). Secondly, the SAID results for $T \leq 100$ MeV are literally swamped by the measurements at higher energies. In case that the floating of the data sets is permitted (as the case is when the 
Arndt-Roper formula is used in the optimization), it is unavoidable that the low-energy behavior of the partial-wave amplitudes will be influenced by the measurements acquired at higher energies. Thirdly, the distribution of 
the normalized residuals, in a self-consistent analysis using a $\chi^2$ minimization function, must be the normal distribution $N(\mu=0,\sigma^2=1)$, where $\mu$ denotes the mean of the distribution and $\sigma^2$ its variance. 
Any effects observed in this distribution (e.g., significant offsets, asymmetry, dependence of the normalized residuals on the independent variables in the problem, etc.) are indicative of problems in the input or in its 
modeling. We are not aware of any paper from the SAID group where these issues are addressed. Regarding the use of the SAID solution for $T \leq 100$ MeV, one final remark is that the measurements from the three $\pi N$ 
reactions, i.e., from the two elastic-scattering (ES) reactions ($\pi^\pm p \to \pi^\pm p$) and from the charge-exchange (CX) reaction ($\pi^- p \to \pi^0 n$), are forced into an isospin-invariant analysis framework. Evidently, 
the SAID group choose to disregard the possibility of the violation of the isospin invariance in the hadronic part of the $\pi N$ interaction~\footnote{In the following, `isospin invariance in the $\pi N$ interaction' will be 
used as the short form of `isospin invariance in the hadronic part of the $\pi N$ interaction'. It is known that the isospin invariance is violated in the electromagnetic (EM) part of the scattering amplitude.} at low energy, 
which has been promulgated by some works during the past two decades \cite{glk,m,mr1}.

\section{\label{sec:Method}Method}

\subsection{\label{sec:KMatrix}Low-energy parameterizations of the hadronic $K$-matrix elements}

The presumptions in analyses employing the $K$-matrix parameterizations of this section relate to: a) the number of terms which one retains from the original infinite power series (expansion of the hadronic $K$-matrix elements 
in terms of a suitable variable, e.g., of the pion kinetic energy $\epsilon$ in the center-of-mass (CM) system), and b) the forms used in the modeling of the resonant contributions. In both cases, we are confident that our 
modeling captures the details of the physical system. Such an approach ensures that the detection of any outliers in the input DBs cannot be attributed to the inability of these parametric forms to account for the energy 
dependence of the hadronic phase shifts; as a result, the detection of outliers in the fits is indicative of experimental discrepancies.

In the analysis of the low-energy $\pi N$ measurements using the $K$-matrix parameterizations, we retain terms up to (and including) $\epsilon^2$. Experience has shown that the coefficients of higher orders in the expansion 
of the $K$-matrix elements cannot be determined from the available measurements at low energy.

\subsubsection{\label{sec:KMPIPEL}Fits to our low-energy $\pi^+ p$ database using the $K$-matrix parameterizations}

For $\pi^+ p$ ES, the $s$-wave phase shift is parameterized as
\begin{equation} \label{eq:S31}
q \cot {\delta}_{0+}^{3/2}=(a_{0+}^{3/2})^{-1} + b_3 \epsilon + c_3 \epsilon^2 \, \, \, ,
\end{equation}
where $q$ denotes (the magnitude of) the CM $3$-momentum. The $p_{1/2}$-wave phase shift is parameterized according to the form
\begin{equation} \label{eq:P31}
\tan {\delta}_{1-}^{3/2}/q = d_{31} \epsilon + e_{31} \epsilon^2 \, \, \, .
\end{equation}

Since the $p_{3/2}$ wave contains the $\Delta (1232)$ resonance, a singular (at $W=m_\Delta$) term must be added to the background term, leading to the expression
\begin{equation} \label{eq:P33}
\tan {\delta}_{1+}^{3/2}/q = d_{33} \epsilon + e_{33} \epsilon^2 + \frac{\Gamma_\Delta m_\Delta}{2 q_\Delta^3 (p_{0 \Delta} + m_p)} \frac{(p_0 + m_p) q^2}{W (m_\Delta-W)} \, \, \, ,
\end{equation}
where $\Gamma_\Delta$ is the $\Delta (1232)$ width, $m_\Delta$ is the $\Delta (1232)$ mass, $m_p$ is the proton mass, $p_0$ is the proton CM energy, and $W$ is the total CM energy. The quantities $q_\Delta$ and $p_{0 \Delta}$ 
denote the values of the variables $q$ and $p_0$, respectively, at the position of the $\Delta (1232)$ resonance ($W=m_\Delta$). The singular term in Eq.~(\ref{eq:P33}) has been obtained from Ref.~\cite{mr2}, see $K_{1+}$ in 
Eqs.~(39) and the corresponding $K^{3/2}_{1+}$ element (after the isospin decomposition of $K_{1+}$ is taken into account), as well as footnote 10 therein.

\subsubsection{\label{sec:KMPIM}Fits to our low-energy $\pi^- p$ elastic-scattering and charge-exchange databases using the $K$-matrix parameterizations}

The isospin $I=3/2$ amplitudes, obtained in the final fit to the truncated low-energy $\pi^+ p$ DB (i.e., to the DB obtained after the removal of the outliers) using the $K$-matrix parameterizations of the preceding section, 
are imported into the analysis of low-energy $\pi^- p$ ES and CX DBs. In this part, another seven parameters (different for these two DBs) are introduced, to parameterize the $I=1/2$ amplitudes. The new parametric forms are 
similar to those given by Eqs.~(\ref{eq:S31}-\ref{eq:P33}), with the parameters $a_{0+}^{1/2}$, $b_1$, $c_1$, $d_{13}$, $e_{13}$, $d_{11}$, and $e_{11}$. Of course, it is necessary to explicitly include the contribution of the 
Roper resonance $N (1440)$ in ${\delta}_{1-}^{1/2}$:
\begin{equation} \label{eq:delta1-1/2}
\tan {\delta}_{1-}^{1/2}/q = d_{11} \epsilon + e_{11} \epsilon^2 + \frac{\Gamma_R M_R (p_{0 R} + m_p)}{2 q_R^3 (M_R+m_p)^2} \frac{(W + m_p)^2 q^2}{W (M_R - W) (p_0 + m_p)} \, \, \, ,
\end{equation}
where (as we are dealing with energies below the pion-production threshold) $\Gamma_R$ is the partial width of the Roper resonance to $\pi N$ decay modes and $m_R$ is its mass. The quantities $q_R$ and $p_{0 R}$ denote the $q$ 
and $p_0$ values at the resonance position ($W=M_R$). The singular term in Eq.~(\ref{eq:delta1-1/2}) has been obtained from Ref.~\cite{mr2}, see Section 3.5.1 therein, in particular, Eq.~(54) for $K_{1-}$.

\subsection{\label{sec:MF}Minimization function}

In our recent partial-wave analyses (PWAs) of the $\pi N$ data, we make use of the minimization function given by the Arndt-Roper formula \cite{ar}, i.e., of the minimization function which the SAID group also use in their 
analyses. The contribution of the $j$-th data set to the overall $\chi^2$ is of the form:
\begin{equation} \label{eq:chijsq}
\chi_j^2=\sum_{i=1}^{N_j} \left( \frac{z_j y_{ij}^{\rm th}-y_{ij}^{\rm exp} }{\delta y_{ij}^{\rm exp} } \right)^2 + \left( \frac{z_j-1}{\delta z_j} \right)^2 \, \, \, ,
\end{equation}
where $y_{ij}^{\rm exp}$ denotes the $i$-th data point of the $j$-th data set, $y_{ij}^{\rm th}$ the corresponding fitted (`theoretical') value, $\delta y_{ij}^{\rm exp}$ the statistical uncertainty of the $y_{ij}^{\rm exp}$ 
data point, $z_j$ a scale factor applied to the entire data set, $\delta z_j$ the normalization uncertainty (reported by the experimental group or assigned by us), and $N_j$ the number of the data points in the data set after 
the removal of any outliers. The fitted values $y_{ij}^{\rm th}$ are obtained by means of the parameterized forms of the $s$- and $p$-wave amplitudes detailed in Sections \ref{sec:KMPIPEL} and \ref{sec:KMPIM}. The values of 
the scale factor $z_j$ are determined (separately for each data set) in such a way as to minimize $\chi_j^2$. For each data set, a unique solution for $z_j$ is obtained via the relation:
\begin{equation} \label{eq:zj}
z_j = \frac{\sum_{i=1}^{N_j} y_{ij}^{\rm th} y_{ij}^{\rm exp} / (\delta y_{ij}^{\rm exp} )^2 + (\delta z_j)^{-2}} {\sum_{i=1}^{N_j} (y_{ij}^{\rm th} / \delta y_{ij}^{\rm exp})^2 + (\delta z_j)^{-2}} \, \, \, ,
\end{equation}
which leads to
\begin{equation} \label{eq:chijsqmin}
(\chi_j^2)_{\rm min} = \sum_{i=1}^{N_j} \frac{ (y_{ij}^{\rm th}-y_{ij}^{\rm exp})^2}{(\delta y_{ij}^{\rm exp})^2 } - \frac {\left( \sum_{i=1}^{N_j}y_{ij}^{\rm th}(y_{ij}^{\rm th}-y_{ij}^{\rm exp}) / (\delta y_{ij}^{\rm exp} )^2 \right)^2} 
{ \sum_{i=1}^{N_j}(y_{ij}^{\rm th}/\delta y_{ij}^{\rm exp})^2 + (\delta z_j)^{-2} } \, \, \, .
\end{equation}
The overall $\chi^2=\sum_{j=1}^{N} (\chi_j^2)_{\rm min}$ (where $N$ denotes the number of the accepted data sets in the fit) is a function of the parameters entering the modeling of the $s$- and $p$-wave amplitudes. These 
parameters are varied until $\chi^2$ attains its minimal value $\chi^2_{\rm min}$.

The part of $(\chi_j^2)_{\rm min}$ which represents the pure random fluctuation in the measurements of the $j$-th data set (i.e., the `unexplained variation' in standard regression terminology) may be obtained from 
Eq.~(\ref{eq:chijsqmin}) in the limit $\delta z_j \to \infty$, which is equivalent to removing the term $(\delta z_j)^{-2}$ from the denominator of the second term on the right-hand side (rhs) of the expression; we denote 
this value by $(\chi_j^2)_{st}$. The variation which is contained in $(\chi_j^2)_{\rm min}$ in excess of $(\chi_j^2)_{st}$ is associated with the contribution from the floating (rescaling) of the data set. The expression for 
$(\chi_j^2)_{sc} \equiv (\chi_j^2)_{\rm min} - (\chi_j^2)_{st}$ is
\begin{equation} \label{eq:chijsqsc}
(\chi_j^2)_{sc} = \frac{(\delta z_j)^{-2} \left( \sum_{i=1}^{N_j}y_{ij}^{\rm th}(y_{ij}^{\rm th}-y_{ij}^{\rm exp}) / (\delta y_{ij}^{\rm exp} )^2 \right)^2}
{\sum_{i=1}^{N_j}(y_{ij}^{\rm th}/\delta y_{ij}^{\rm exp})^2 \left( \sum_{i=1}^{N_j}(y_{ij}^{\rm th}/\delta y_{ij}^{\rm exp})^2 + (\delta z_j)^{-2} \right) } \, \, \, .
\end{equation}
The scale factors which minimize only the first term on the rhs of Eq.~(\ref{eq:chijsq}) are obtained from Eq.~(\ref{eq:zj}) in the limit $\delta z_j \to \infty$:
\begin{equation} \label{eq:zjopt}
\hat{z}_j = \frac{\sum_{i=1}^{N_j} y_{ij}^{\rm th} y_{ij}^{\rm exp} / (\delta y_{ij}^{\rm exp} )^2 } {\sum_{i=1}^{N_j} (y_{ij}^{\rm th} / \delta y_{ij}^{\rm exp})^2 } \, \, \, .
\end{equation}
The scale factors $\hat{z}_j$ represent the optimal floating of the fitted values $y_{ij}^{\rm th}$ around the experimental results $y_{ij}^{\rm exp}$, with no regard to the normalization uncertainty.

The analysis of this work may be performed using either of the two scale factors, $z_j$ or $\hat{z}_j$. We make use of the scale factors $z_j$, as only these quantities are contained in the output of the SAID Analysis Program.

For the purpose of the optimization, we employ the standard MINUIT package \cite{jms} of the CERN library (FORTRAN version). Each optimization has been achieved on the basis of the (robust) SIMPLEX-MINIMIZE-MIGRAD-MINOS 
sequence. All fits of this work terminated successfully.

\section{\label{sec:KM}Results from our partial-wave analysis}

The list of the experiments, which are included in our three $\pi N$ DBs, is available from Refs.~\cite{mr1,mr3}; the same notation will be used here to identify the individual data sets. Our DBs contain differential cross 
sections (DCSs), analyzing powers (APs), partial-total cross sections (PTCSs), and total (as well as total-nuclear) cross sections (TCSs) for $T \leq 100$ MeV. The largest difference between the SAID DB for $T \leq 100$ MeV 
and ours relates to the ES DCSs of the CHAOS Collaboration \cite{chaos,denz}. Regarding these measurements, our opinion is known \cite{mr4,mr5} and there is no reason to repeat it here.

Modifications in the analysis software and DB structure in the recent years enable us now to also include in the optimization the APs of Ref.~\cite{meier}, comprising a total of $28$ data points. There had been two technical 
reasons preventing the direct use of these measurements in our PWAs before 2013: a) each of the three data sets, to which the measurements of Ref.~\cite{meier} must be assigned, involves more beam energies than one and b) the 
last of these data sets contains measurements of both ES reactions.

Regarding the proton EM form factors, recent developments suggest the replacement of the forms we had been using before 2013. The parameterization of the Dirac $F_1^p (t)$ and Pauli $F_2^p (t)$ form factors of the proton with 
(traditional) dipole forms has been found to provide a poor description of the `world' electron-proton ($e p$) unpolarized and polarized data \cite{bern}. Although the sensitivity of our results to the details of the 
parameterization of these quantities is low (due to the smallness of the $Q^2$ transfer for $T \leq 100$ MeV), adopted now is an improved parameterization. In Ref.~\cite{vamx}, the authors made use of the so-called Pad\'e 
parameterization \cite{amt} for the Sachs EM form factors $G_E^p (t)$ and $G_M^p (t)$ (the superscript $p$ is omitted in Ref.~\cite{vamx}), and obtained the optimal values of the relevant parameters from a fit to $e p$ 
measurements; we now use the results of their Table II. The pion form factor $F^\pi (t)$ is usually parameterized via a monopole form, e.g., see Ref.~\cite{na7}. Although results of the same quality are obtained in the 
low-energy region with either a monopole or a dipole form, the monopole parameterization is now adopted.

The values of the relevant physical constants (see Table \ref{tab:Constants}) have been fixed from the 2014 compilation of the Particle-Data Group (PDG) \cite{pdg}. Finally, an improved approach for determining the 
(small) $d$ and $f$ waves has been implemented; to suppress artefacts which are due to the truncation of small values, simple polynomials are now fitted to the $d$- and $f$-wave phase shifts of the SAID analysis \cite{abws}, 
which (as it incorporates dispersion-relation constraints) is expected to determine reliably these phase shifts in the region $T > 100$ MeV.

One statistical test for each data set is performed, the one involving its contribution $(\chi_j^2)_{\rm min}$ to the overall $\chi^2_{\rm min}$ (see Section \ref{sec:MF}). The p-value, estimated from $(\chi_j^2)_{\rm min}$ 
and $N_j$, is compared to the user-defined confidence level $\mathrm{p}_{\rm min}$ for the acceptance of the null hypothesis (no statistically significant effects); in case that the extracted p-value is below $\mathrm{p}_{\rm min}$, 
the degree of freedom (DOF) with the largest contribution to $(\chi_j^2)_{\rm min}$ is eliminated in the subsequent fit. As in our recent PWAs, we adopt the $\mathrm{p}_{\rm min}$ value which is associated with a $2.5 \sigma$ 
effect in the normal distribution. This value is approximately equal to $1.24 \cdot 10^{-2}$, i.e., slightly larger than $1.00 \cdot 10^{-2}$, a threshold which most statisticians recommend as the outset of statistical 
significance.

When identifying the outliers in our past PWAs, the maximal number of excluded data points for each data set had been fixed (somewhat arbitrarily) at $2$; data sets with more excluded DOFs were removed from the analysis. To 
do justice to data sets containing a large number of data points, this restriction was recently revised~\footnote{This modification does not affect any of the results we have obtained thus far for $\mathrm{p}_{\rm min} \approx 1.24 \cdot 10^{-2}$.}. 
Assuming pure statistical fluctuation, the probability that the $j$-th data set contains at most $2$ outliers decreases with increasing $N_j$ values. Evidently, the maximal number of outliers must be determined separately for 
each data set, on the basis of the number of measurements which the data set initially contains.

Let us assume that the \emph{a-priori} probability for a data point to be an outlier is equal to p. The probability that the $j$-th data set (with $N_j$ initial data points) contains exactly $k$ outliers is then given by the 
expression
\begin{equation} \label{eq:Exc}
P_k = \binom{N_j}{k} \mathrm{p}^k (1-\mathrm{p})^{N_j-k} \, \, \, .
\end{equation}
Consequently, the probability that the $j$-th data set contains up to $N_j^{\rm out} \leq N_j$ outliers is given by the sum
\begin{equation} \label{eq:Exc1}
P(N_j^{\rm out}) = \sum_{k=0}^{N_j^{\rm out}} P_k \, \, \, .
\end{equation}
After $\mathrm{p}_{\rm min}$ is set, the maximal number of outliers permitted for the particular experiment (at that $\mathrm{p}_{\rm min}$ level) may be identified as the maximal $N_j^{\rm out}$ value for which the cumulative 
probability, obtained with Eq.~(\ref{eq:Exc1}), does not exceed $1-\mathrm{p}_{\rm min}$. It thus follows that, if $\mathrm{p}=0.06$, a data set with $20$ data points will be allowed to contain $3$ outliers, whereas one with 
only $10$ data points a maximum of $2$ outliers for the $\mathrm{p}_{\rm min}$ value adopted herein. To obtain an estimate of p, the maximal number of outliers $N_j^{\rm out}$ was set equal to $N_j$ (which is equivalent to 
relaxing the condition for removing entire data sets) and the fits to our low-energy $\pi^+ p$ DB (which is known to contain the largest proportion of outliers) were carried out iteratively, excluding at each iteration step 
the data point with the largest contribution to the overall $\chi^2_{\rm min}$. The procedure was repeated until no data point could be identified as an outlier. An estimate of the probability p is obtained as the fraction of 
the number of DOFs (NDF) of the initial $\pi^+ p$ DB which had to be eliminated. To avoid the exclusion of low-$N_j$ data sets just because of the removal of $1$ DOF, $N_j^{\rm out}$ was finally redefined as $\max \{ N_j^{\rm out}, 2\}$.

Data sets which do not give acceptable p-values (i.e., exceeding $\mathrm{p}_{\rm min}$) after the elimination of the appropriate number of data points (the absolute normalization is also considered to be one of the acquired 
measurements), as explained above, were removed from the DB. Only one point was removed at each step. The optimization was repeated, until no data point could be identified as an outlier. It must be emphasized that, in our 
approach, the identification of the outliers in each of the three reactions is based on comparisons involving \emph{only} the measurements in that particular reaction.

\subsection{\label{sec:AnalPIPEL}Fits to our low-energy $\pi^+ p$ database}

Our initial low-energy $\pi^+ p$ DB contains $57$ data sets consisting of $389$ data points. As in all our past PWAs, we found that the data sets of BRACK90 at $66.80$ MeV and JORAM95 at $32.70$ MeV (with $11$ and $7$ data 
points, respectively) had to be removed from the DB. Subsequently, three data sets had to be freely floated, namely two of the BRACK86 data sets (at $66.80$ and $86.80$ MeV), as well as the BRACK90 data set at $30.00$ MeV. 
Three additional single data points had to be removed. After the elimination of $24$ DOFs of the initial DB, we obtained a truncated low-energy $\pi^+ p$ DB comprising $55$ data sets and $365$ DOFs. The accepted data sets are 
detailed in Table \ref{tab:DBPIPEL}.

The $\chi^2_{\rm min}$ value, corresponding to the fit to the initial DB, was $706.7$; for the truncated DB, $\chi^2_{\rm min} \approx 459.5$. Therefore, the removal of $24$ DOFs from the initial DB results in the decrease of 
the $\chi^2$ by $247.2$ units, corresponding to more than $10$ units per removed entry on average. At the same time, the p-value of the fit increased by over $17$ orders of magnitude.

The scale factors $z_j$, corresponding to the DCSs in our low-energy $\pi^+ p$ DB~\footnote{Being ratios of cross sections, the APs are not suitable for the demonstration of the systematic effects which we have set about 
investigating in this study. The PTCSs and the TCSs (all one- or two-point data sets) have been used in the optimization, but have not been included in the analysis of the scale factors shown in Fig.~\ref{fig:PIPELKMatrix}, as 
the SAID group do not include these measurements in their DB.}, are shown in Fig.~\ref{fig:PIPELKMatrix}. The weighted linear fit to the data shown (the weight of each entry is equal to $(\delta z_j)^{-2}$) yields an intercept 
of $1.023 \pm 0.033$ and a slope of $(-0.38 \pm 0.40) \cdot 10^{-3}$ MeV$^{-1}$; both fitted values are compatible with the expectation values (of $1$ and $0$, respectively) for a successful optimization.

It is sometimes suggested that the joint analysis of the DCSs and the PTCSs in the $\pi^+ p$ reaction may be inappropriate, as the integrated DCS might not match the measured PTCS. Such a mismatch could be due to the effects 
of a presently unknown state characterized by a $\pi N$ decay mode, e.g., as the case would be if a $\pi N$ resonance, not currently established, existed below the $\Delta(1232)$ mass (see also Section \ref{sec:Conclusions}). 
We have investigated the sensitivity of our results to the inclusion of the PTCSs/TCSs in the fitted $\pi^+ p$ DB and have concluded that our analysis is very little affected by the treatment of these measurements. To demonstrate 
this, we also mention the fitted values of the intercept and of the slope after the $\pi^+ p$ PTCSs/TCSs are removed from the input DB: the new intercept comes out equal to $1.025 \pm 0.028$, whereas the new slope is 
$(-0.31 \pm 0.36) \cdot 10^{-3}$ MeV$^{-1}$. We performed this check only for the sake of completeness, with no intention to cast doubt on the validity of the low-energy $\pi^+ p$ PTCSs/TCSs. In our approach, we have found no 
evidence of mismatch between the DCSs and the PTCSs/TCSs: these two observables can be accounted for in a joint optimisation scheme.

\subsection{\label{sec:AnalPIMEL}Fits to our low-energy $\pi^- p$ elastic-scattering database}

Our initial low-energy $\pi^- p$ ES DB contains $37$ data sets consisting of $339$ data points. The BRACK90 data set at $66.80$ MeV (with $5$ data points in total) needed to be removed, one data set (the WIEDNER89 data set at 
$54.30$ MeV) had to be freely floated, and two additional single data points had to be removed. After the elimination of $8$ DOFs of the initial DB, we obtained a truncated low-energy $\pi^- p$ ES DB comprising $36$ data sets 
and $331$ DOFs. The accepted data sets are detailed in Table \ref{tab:DBPIMEL}.

The $\chi^2_{\rm min}$ value, corresponding to the fit to the initial DB, was $524.8$; for the truncated DB, $\chi^2_{\rm min} \approx 371.0$. Therefore, the removal of only $8$ DOFs from the initial DB results in a decrease 
of the $\chi^2$ by $153.8$ units, i.e., just short of $20$ units per removed entry on average. At the same time, the p-value of the fit increased by over $8$ orders of magnitude.

The scale factors $z_j$, corresponding to the DCSs in our low-energy $\pi^- p$ ES DB, are shown in Fig.~\ref{fig:PIMELKMatrix}. The weighted linear fit to the data shown yields an intercept of $1.007 \pm 0.019$ and a slope of 
$(-0.08 \pm 0.24) \cdot 10^{-3}$ MeV$^{-1}$; both fitted values are compatible with the expectation for a successful optimization.

\subsection{\label{sec:AnalPIMCX}Fits to our low-energy $\pi^- p$ charge-exchange database}

Our initial low-energy $\pi^- p$ CX DB contains $54$ data sets consisting of $333$ data points. Only $5$ DOFs needed to be removed: the FITZGERALD86 data sets at $32.48$, $36.11$, $40.26$, and $47.93$ MeV needed to be freely 
floated, and the BREITSCHOPF06 TCS at $75.10$ MeV had to be removed. After the elimination of these $5$ DOFs of the initial DB, we obtained a truncated low-energy $\pi^- p$ CX DB comprising $53$ data sets and $328$ DOFs. The 
accepted data sets are detailed in Table \ref{tab:DBPIMCX}.

The $\chi^2_{\rm min}$ value, corresponding to the fit to the initial DB, was $401.6$; for the truncated DB, $\chi^2_{\rm min} \approx 313.7$. Therefore, the removal of only $5$ DOFs from the initial DB results in a decrease 
of the $\chi^2$ by $87.9$ units, i.e., to $17.6$ units per removed entry on average.

The scale factors $z_j$, corresponding to the DCSs in our low-energy $\pi^- p$ CX DB, are shown in Fig.~\ref{fig:PIMCXKMatrix}. The weighted linear fit to the data shown yields an intercept of $1.002 \pm 0.015$ and a slope of 
$(0.05 \pm 0.27) \cdot 10^{-3}$ MeV$^{-1}$; again, both fitted values are compatible with the expectation for a successful optimization.

\section{\label{sec:WI08}Results from the WI08 solution}

The SAID $\pi^+ p$ DB for $T \leq 100$ MeV contains $721$ data points, $679$ of which relate to DCSs, the remaining $42$ to APs; PTCSs and TCSs are not included. The DB contains the entirety of the DENZ05 data, as well as the 
BERTIN76 measurements (save for the $67.40$ MeV data set). The BERTIN76, the AULD79, and the FRANK83 data sets are floated~\footnote{The definition of floating for the SAID group is not identical to ours. For us, free floating 
involves the limit $\delta z_j \to \infty$, whereas they simply set the corresponding normalization uncertainty $\delta z_j$ of the data set to the value of $1$. This value drastically reduces, but does not eliminate, the 
contribution of the floating contribution $(\chi_j^2)_{sc}$ of Eq.~(\ref{eq:chijsqsc}) to $(\chi_j^2)_{\rm min}$ of Eq.~(\ref{eq:chijsqmin}).}. The overall $\chi^2_{\rm min}$, corresponding to the SAID $\pi^+ p$ DB for $T \leq 100$ 
MeV, is equal to $1595.2$, yielding a p-value of a few $10^{-68}$. Evidently, the description of the SAID $\pi^+ p$ DB for $T \leq 100$ MeV with the WI08 phase-shift solution is poor.

The SAID $\pi^- p$ ES DB for $T \leq 100$ MeV contains $634$ data points, $546$ of which relate to DCSs, the remaining $88$ to APs. Similar to us (but for different reasons~\footnote{Given that the nine available $\pi^- p$ 
PTCSs and TCSs contain a component from CX scattering, these measurements have never been included in our $\pi^- p$ ES DB. Their involvement in any part of the analysis (see beginning of Section \ref{sec:SimilaritiesAndDifferences}) 
would complicate the discussion on the violation of the isospin invariance in the $\pi N$ interaction.}), they have not included the corresponding PTCSs and TCSs in their DB. Their DB contains the entirety of the DENZ05 data. 
The four FRANK83 data sets are floated. The overall $\chi^2_{\rm min}$, corresponding to the SAID $\pi^- p$ ES DB for $T \leq 100$ MeV, is equal to $1102.1$, yielding a p-value of about $10^{-27}$. Therefore, the description 
of the SAID $\pi^- p$ ES DB for $T \leq 100$ MeV with the WI08 solution is almost as poor as that of their $\pi^+ p$ DB.

The SAID $\pi^- p$ CX DB for $T \leq 100$ MeV contains $353$ data points, $343$ of which relate to DCSs, the remaining $10$ to APs; the corresponding TCSs are not included therein. The three DUCLOS73 measurements have been 
deleted from their DB. The overall $\chi^2_{\rm min}$, corresponding to the SAID $\pi^- p$ CX DB for $T \leq 100$ MeV, is equal to $401.3$, yielding a p-value of about $3.9 \cdot 10^{-2}$. The description of the SAID $\pi^- p$ 
CX DB for $T \leq 100$ MeV with the WI08 solution is acceptable.

There are three reasons why the description of \emph{their} low-energy ES DBs with \emph{their} WI08 solution is poor.
\begin{itemize}
\item The SAID $\pi N$ DB is very extensive. The low-energy behavior of the partial-wave amplitudes is largely determined from measurements acquired at higher energies. As the total amount of the measurements, contained in 
their DB, exceeds $50\,000$ data points, their $1\,708$ low-energy measurements are literally swamped in the analysis.
\item The SAID group include in their DBs all the available measurements (including a few data sets which have not appeared in formal publications); outliers are seldom excluded.
\item The three $\pi N$ reactions are subject to an analysis which assumes the fulfillment of the isospin invariance in the $\pi N$ interaction at all energies.
\end{itemize}

We will now report the results of a very simple analysis of the scale factors $z_j$ for $T \leq 100$ MeV, as these quantities come out of their fits to the data. These values (two decimal digits are available online) were 
obtained from the SAID web page on May 6, 2016; we also use their $T_j$ and $\delta z_j$ values. Scatter plots of the scale factors $z_j$ versus the corresponding beam energy $T_j$ for the data sets of the SAID DB are shown, 
\emph{separately} for the three $\pi N$ reactions, in Figs.~\ref{fig:PIPELWI08}-\ref{fig:PIMCXWI08}. The fitted values of the parameters of the weighted linear fits (as well as their uncertainties, corrected with the Birge 
factor $\sqrt{\chi^2/{\rm NDF}}$), are shown in Table \ref{tab:WI08Parameters}. Visual inspection of these figures and of Table \ref{tab:WI08Parameters} leads to two conclusions:
\begin{itemize}
\item In all three $\pi N$ reactions, the departure of the scale factors $z_j$ from the expectation for a successful optimization is noticeable. The energy dependence of the scale factors $z_j$ is more pronounced in the case 
of the $\pi^+ p$ reaction.
\item The departure from the expectations appears to occur slightly below $T=100$ MeV; decreasing effects with increasing beam energy are observed.
\end{itemize}

If the scale factors from the three $\pi N$ reactions are analyzed in a joint scheme, then the systematic effects, observed in the individual reactions, disappear. This becomes evident after comparing the last row of Table 
\ref{tab:WI08Parameters} with the previous three, as well as after comparing Fig.~\ref{fig:ALLWI08} with Figs.~\ref{fig:PIPELWI08}-\ref{fig:PIMCXWI08}.

\section{\label{sec:SimilaritiesAndDifferences}Similarities and differences between the two approaches}

In our method of extracting the important information from the $\pi N$ measurements for $T \leq 100$ MeV, three steps are followed. At the first step, we employ simple parameterizations of the $s$- and $p$-wave $K$-matrix 
elements, retaining in the expressions orders up to $\epsilon^2$. We make use of these general parameterizations in order to reliably identify and remove any outliers present in the initial DBs, thus obtaining self-consistent 
input for the subsequent phases of the analysis. At the second step of our approach, the ETH model of the $\pi N$ interaction (a complete description of this model and details on its development may be found in Ref.~\cite{mr2}) 
is fitted to the truncated DBs of the two ES reactions, thus leading to the determination of the values of the model parameters (coupling constants and vertex factors), which account optimally for the measurements of these two 
reactions. At a third step, we investigate the violation of the isospin invariance in the $\pi N$ interaction, by comparing the model predictions for the CX reaction (obtained from the fitted values, as well as the correlation 
matrix of the fit to the two ES DBs) to the measurements of the CX reaction. We reported in the past that these predictions significantly \emph{underestimate} the CX DCSs in most of the low-energy region. Assuming the correctness 
of the bulk of the available measurements and the smallness of any missing pieces in the EM corrections, this mismatch between predictions and measurements strongly indicates the violation of the isospin invariance in the 
$\pi N$ interaction at low energy (for a more detailed discussion, see Section 7 of Ref.~\cite{mr1}). Following a slightly different methodology (and fewer CX measurements), Gibbs, Ai, and Kaufmann came to the same conclusion 
in the mid 1990s \cite{glk}.

Our PWAs are restricted to $T \leq 100$ MeV, because the ETH model is expected to work better in the low-energy region and because our EM corrections \cite{gmorw1,gmorw2} have been established only below $100$ MeV. Regarding 
the former remark, the introduction of strong-interaction form factors in the Feynman graphs of the ETH model is unnecessary below $100$ MeV and the contributions from graphs involving distant baryonic states - i.e., the higher 
baryon resonances with masses above $2$ GeV - to the $s$- and $p$-wave amplitudes of the model are expected to be negligible.

A dispersion-relation analysis framework, such as the one developed by the SAID group, assumes the fulfillment of the isospin invariance in the $\pi N$ interaction and analyzes the entire DB available, extending to $T$ values 
of several GeV, so that the dispersion integrals be evaluated reliably. Such an approach cannot be performed in a restricted energy region, without bringing in external influences. Such a framework cannot be used in order to 
conduct exclusive analyses of the data, be they restricted in energy or involving only one reaction. This is one important difference between our approach and theirs. The second difference relates to the procedure: we choose 
to analyze the two ES processes (this analysis fixes the $I=3/2$ and $I=1/2$ amplitudes) and subsequently investigate whether isospin invariance is fulfilled in the $\pi N$ interaction, on the basis of comparisons with the 
measurements of the third reaction (triangle identity~\footnote{Assuming that the isospin invariance holds in the $\pi N$ interaction, only two (of course, complex) amplitudes enter the physical description of the three $\pi N$ 
reactions: the $I=3/2$ amplitude ($f_3$) and the $I=1/2$ amplitude ($f_1$). Disregarding the distortions which are due to the EM effects, the $\pi^+ p$ reaction is described by $f_3$, the $\pi^- p$ ES reaction by $(2 f_1 + f_3)/3$, 
and the CX reaction by $\sqrt{2} (f_3-f_1)/3$. On the basis of these relations, the triangle identity between the three corresponding amplitudes $f_{\pi^+ p}$, $f_{\pi^- p}$, and $f_{CX}$ is obtained: 
$f_{\pi^+ p} - f_{\pi^- p} = \sqrt{2} f_{CX}$.}). On the contrary, the SAID group assume the fulfillment of the isospin invariance in the $\pi N$ interaction at all energies, and analyze the three $\pi N$ reactions in a joint 
optimization scheme.

So far, the emphasis has been placed on the differences between the two approaches. However, there are also similarities. For instance, both approaches rely on the correctness of the input data. To an extent, we take steps 
towards investigating the self-consistency of the input DB, whereas they avoid excluding measurements. Only considering the number of the input data points, the sensitivity of our analysis to outliers is expected to be more 
pronounced than it is in their case: $500$ discrepant DCSs in a DB of $1\,000$ data points wreck havoc; the same amount of discrepant measurements in a DB of over $50\,000$ data points is surely less problematic. As a result, 
we cannot but thoroughly examine every new data set prior to including it in the DB; in this respect, they can afford to be more generous. Finally, both approaches rely on the correctness/completeness of the method incorporating 
the EM effects (distortion corrections to the hadronic phase shifts and to the partial-wave amplitudes) in the analysis.

To summarize, the SAID group assume that isospin invariance in the $\pi N$ interaction is fulfilled at all energies, whereas we test whether it is at low energy. They cannot easily restrict their analysis to specific energy 
regions or reactions, whereas we may perform a variety of analyses in relation to the input DBs: we may perform joint analyses (e.g., by using as input the $\pi^+ p$ DB along with either of the $\pi^- p$ DBs or by submitting 
the measurements of all three $\pi N$ reactions to a joint optimization scheme) or we may analyze the three $\pi N$ reactions independently of one another (using our $K$-matrix parameterizations). However, our analyses can 
only involve the low-energy region. Both approaches rely on the correctness of the bulk of the measurements (ours is significantly more sensitive to the presence of outliers in the DB) and both approaches rely on the correctness 
of the EM corrections, applied to the hadronic phase shifts and to the partial-wave amplitudes on the way to fitting the observables.

\section{\label{sec:Conclusions}Discussion and conclusions}

\begin{figure}
\begin{center}
\includegraphics [width=15.5cm] {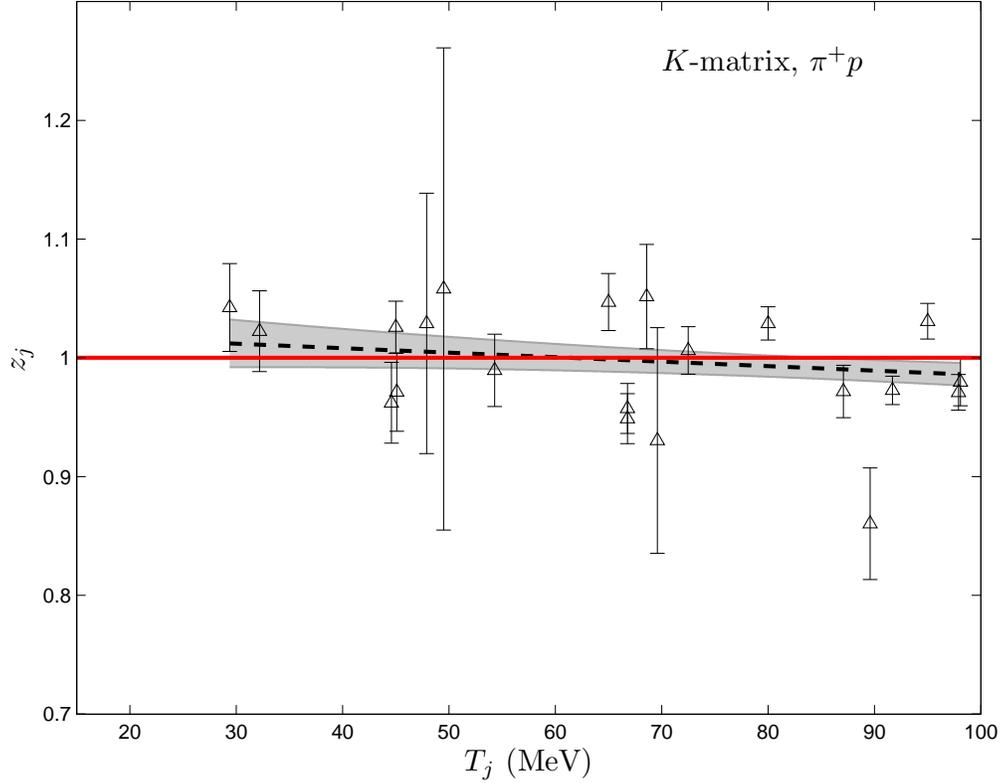}
\caption{\label{fig:PIPELKMatrix}Plot of the scale factors $z_j$ which minimize $\chi_j^2$ of Eq.~(\ref{eq:chijsq}) for the fits to our low-energy $\pi^+ p$ database using the $K$-matrix parameterizations of Section \ref{sec:KMPIPEL}; 
$T_j$ denotes the pion laboratory kinetic energy of the $j$-th experiment. The data correspond to DCSs only; the data sets which had to be freely floated are not shown. The dashed straight line represents the optimal, weighted 
linear fit to the data shown and the shaded band $1 \sigma$ uncertainties around the fitted values. The red line is the optimal, unbiased outcome of the optimization.}
\end{center}
\end{figure}

\begin{figure}
\begin{center}
\includegraphics [width=15.5cm] {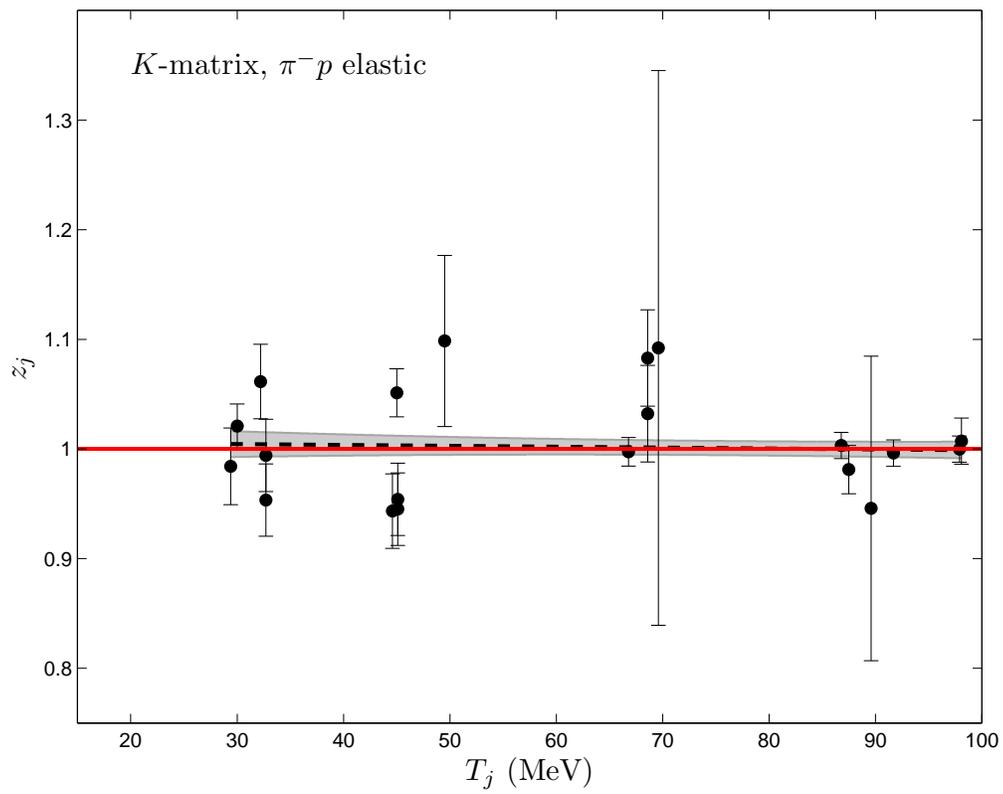}
\caption{\label{fig:PIMELKMatrix}The equivalent of Fig.~\ref{fig:PIPELKMatrix} for the fits to our $\pi^- p$ elastic-scattering database.}
\end{center}
\end{figure}

\begin{figure}
\begin{center}
\includegraphics [width=15.5cm] {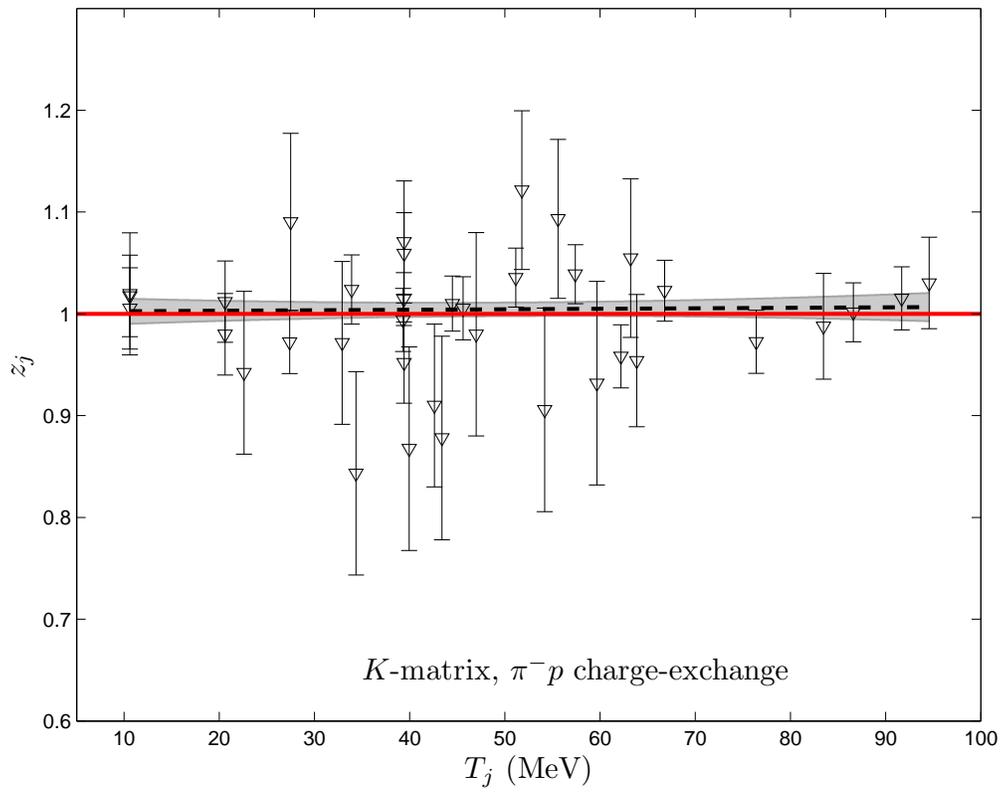}
\caption{\label{fig:PIMCXKMatrix}The equivalent of Fig.~\ref{fig:PIPELKMatrix} for the fits to our $\pi^- p$ charge-exchange database.}
\end{center}
\end{figure}

A self-consistent, $\chi^2$-based optimization scheme satisfies the requirement that the distribution of the normalized residuals of the fit follow the $N(0,1)$ distribution. Furthermore, these residuals must not exhibit 
significant dependence on the independent variables in the problem, e.g., on the beam energy in this work. In a self-consistent optimization scheme, where the input measurements are reliable and their modeling adequate, all 
fluctuations present in the residuals are random. We demonstrated that the results of the weighted linear fits to the scale factors $z_j$, obtained from the data when using our $K$-matrix parameterizations, come out as expected: 
independent of the beam energy and `clustering' around the expectation value of $1$ (see Figs.~\ref{fig:PIPELKMatrix}-\ref{fig:PIMCXKMatrix}).

The SAID analysis rests upon the fulfillment of three conditions \emph{at all energies}:
\begin{itemize}
\item[(a)] that the bulk of the experimental data is reliable;
\item[(b)] that the EM effects are correctly accounted for; and
\item[(c)] that the isospin invariance is fulfilled in the $\pi N$ interaction.
\end{itemize}

When the Arndt-Roper formula is used in the optimization, one expects that the data sets which must be scaled `upwards' should be balanced (on average) by those which must be scaled `downwards'. Additionally, the energy 
dependence of the scale factors must not be significant. If these prerequisites are not fulfilled, the modeling of the input measurements cannot be considered to be satisfactory. The scale factors, relating to the description 
of the SAID low-energy $\pi N$ DBs with the WI08 solution, are shown in Fig.~\ref{fig:ALLWI08}. At first glance, the plot leaves a satisfactory impression, in particular after considering the extent of the SAID $\pi N$ DB, as 
well as the practice of the SAID group to avoid excluding the outliers. Consequently, the optimization scheme leading to the WI08 solution yields normalized residuals for the scale factors~\footnote{These residuals are equal 
to $(z_j-1)/\delta z_j$, see Eq.~(\ref{eq:chijsq}).} which are reasonably well centered on $0$ (we disregard small effects). Given that the reduced $\chi^2$, corresponding to the entirety of the fitted data at low energy in 
the case of the WI08 solution, is about $1.81$, the joint analysis of the data inevitably yields normalized residuals with a distribution broader than the $N(0,1)$ distribution. However, this broadening originates in the choice 
of the SAID group to seldom exclude outliers.

\begin{figure}
\begin{center}
\includegraphics [width=15.5cm] {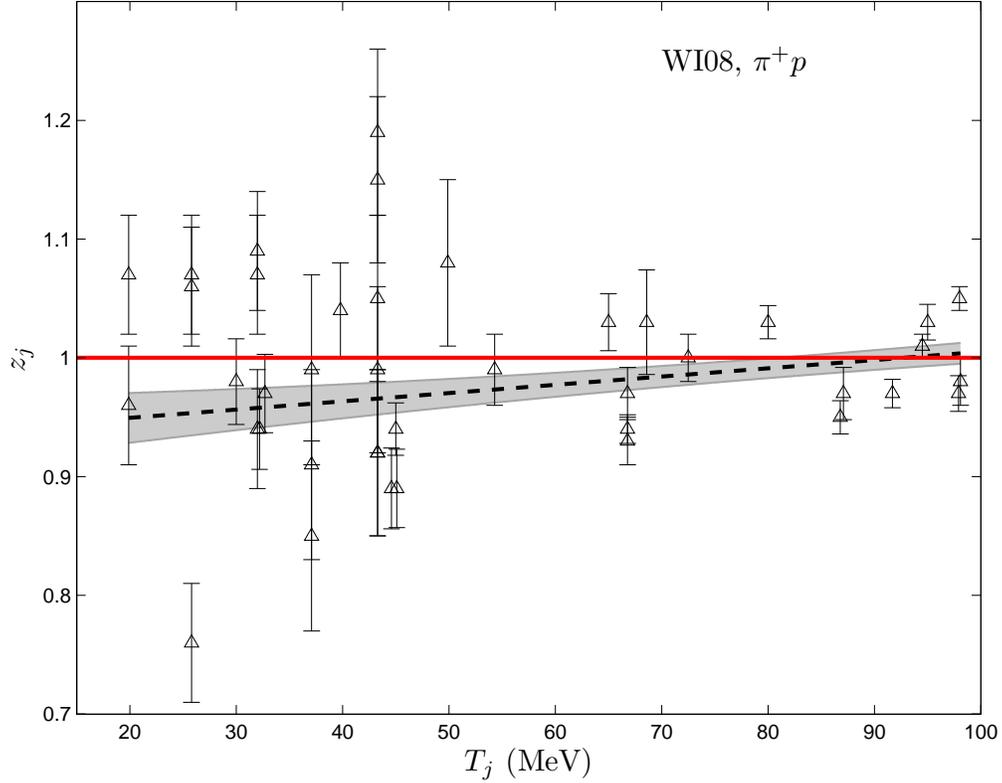}
\caption{\label{fig:PIPELWI08}Plot of the scale factors $z_j$ which minimize $\chi_j^2$ of Eq.~(\ref{eq:chijsq}) for the fits to the SAID low-energy $\pi^+ p$ database (yielding the WI08 solution); $T_j$ denotes the pion 
laboratory kinetic energy of the $j$-th experiment. The data correspond to DCSs only; the data sets which were floated are not shown. The dashed straight line represents the optimal, weighted linear fit to the data shown (see 
Table \ref{tab:WI08Parameters}) and the shaded band $1 \sigma$ uncertainties around the fitted values. The red line is the optimal, unbiased outcome of the optimization.}
\end{center}
\end{figure}

\begin{figure}
\begin{center}
\includegraphics [width=15.5cm] {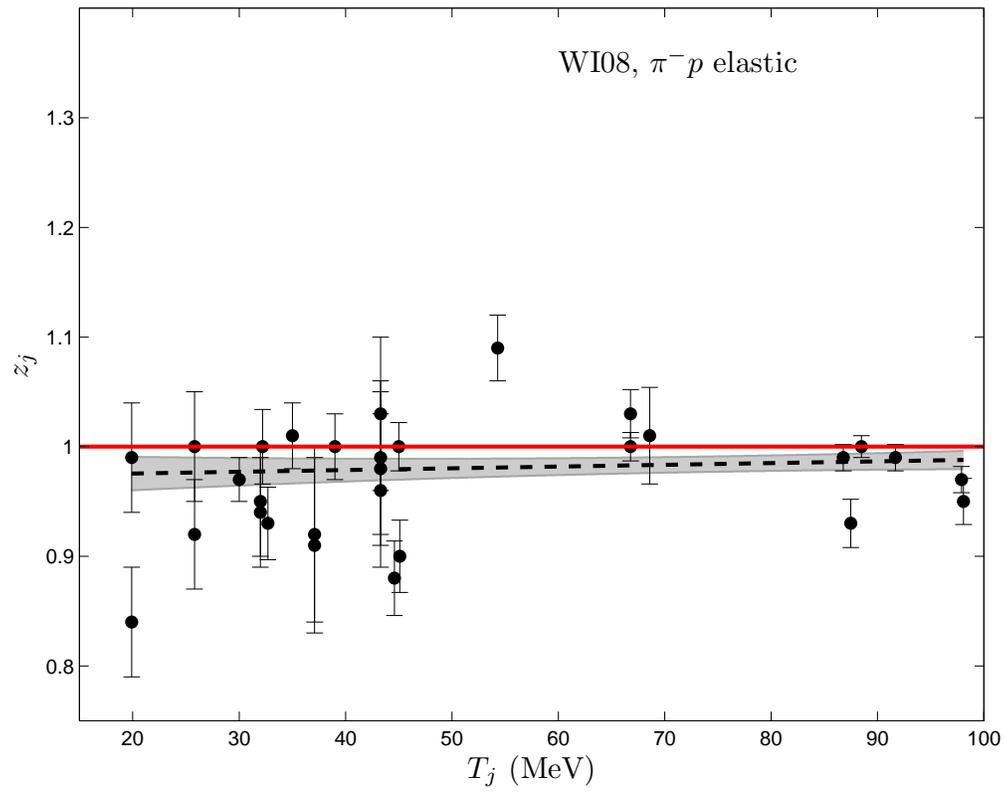}
\caption{\label{fig:PIMELWI08}The equivalent of Fig.~\ref{fig:PIPELWI08} for the SAID low-energy $\pi^- p$ elastic-scattering database.}
\end{center}
\end{figure}

\begin{figure}
\begin{center}
\includegraphics [width=15.5cm] {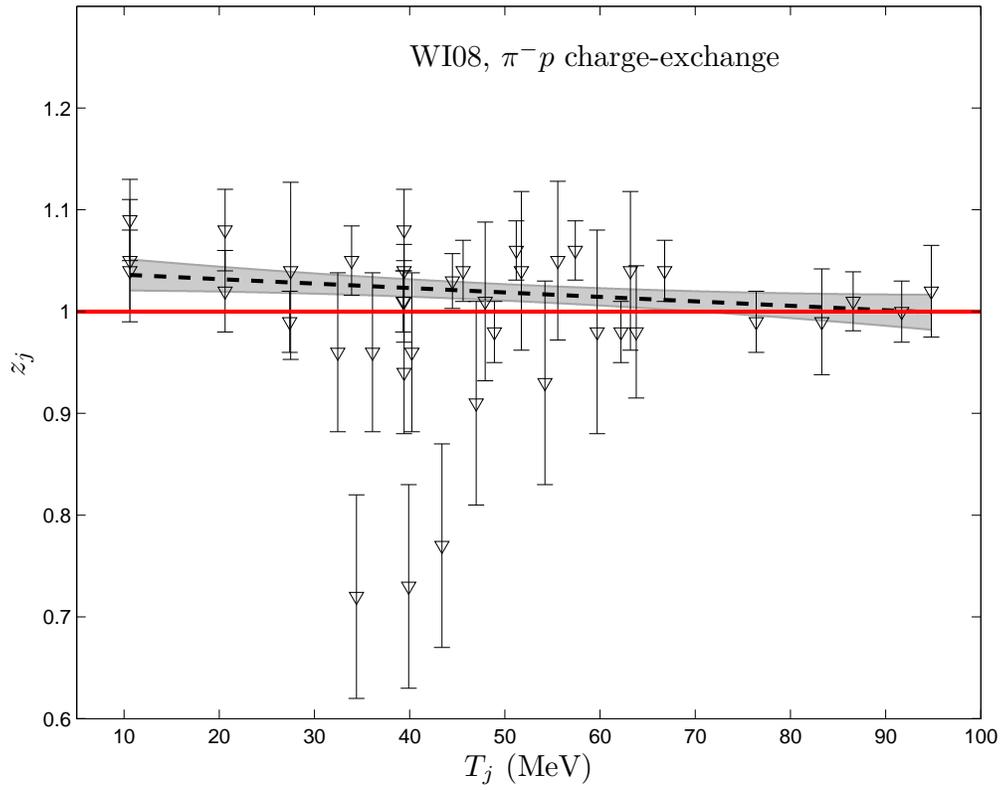}
\caption{\label{fig:PIMCXWI08}The equivalent of Fig.~\ref{fig:PIPELWI08} for the SAID low-energy $\pi^- p$ charge-exchange database.}
\end{center}
\end{figure}

\begin{figure}
\begin{center}
\includegraphics [width=15.5cm] {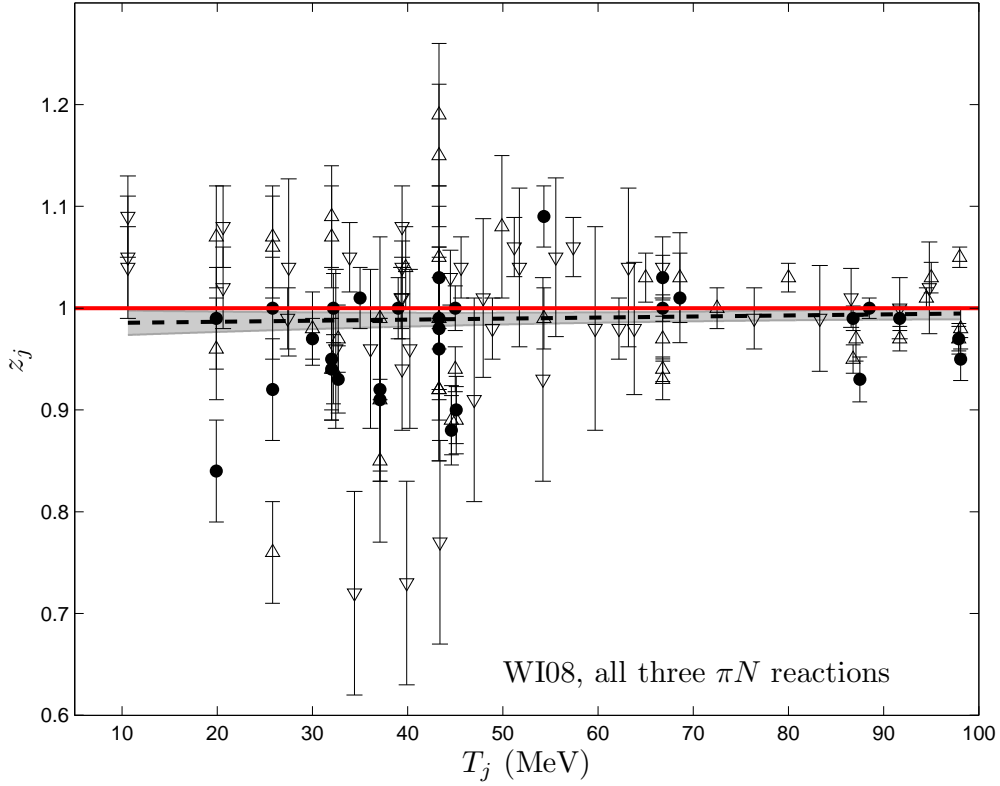}
\caption{\label{fig:ALLWI08}Plot of the scale factors $z_j$ which minimize $\chi_j^2$ of Eq.~(\ref{eq:chijsq}) for the fits to the entire SAID low-energy $\pi N$ database (yielding the WI08 solution); $T_j$ denotes the pion 
laboratory kinetic energy of the $j$-th experiment. The data correspond to DCSs only (upward triangles: $\pi^+ p$, dots: $\pi^- p$ elastic, downward triangles: $\pi^- p$ charge-exchange); the data sets which were floated are 
not shown. The dashed straight line represents the optimal, weighted linear fit to the data shown (see Table \ref{tab:WI08Parameters}) and the shaded band $1 \sigma$ uncertainties around the fitted values. The red line is the 
optimal, unbiased outcome of the optimization.}
\end{center}
\end{figure}

Let us finally express our criticism on the low-energy behavior of the WI08 solution. The SAID input DB consists of three distinct parts, identified as the sets of measurements of the three $\pi N$ reactions. Had their fit been 
unbiased, the general behavior of the fitted $z_j$ values in terms of their dependence on $T_j$, as obtained from Fig.~\ref{fig:ALLWI08}, would also have been observed in any arbitrary subset of their DB, consistent with the 
basic principles of the Sampling Theory (adequate population, representative sampling). The scale factors, relating to the description of the SAID low-energy $\pi N$ DBs with the WI08 solution, are shown (separately for the 
three reactions) in Figs.~\ref{fig:PIPELWI08}-\ref{fig:PIMCXWI08}. Had the three aforementioned conditions (a)-(c) been fulfilled in the SAID analysis at all energies, the scale factors of Figs.~\ref{fig:PIPELWI08}-\ref{fig:PIMCXWI08} 
would have come out independent of the beam energy and would have been centered on $1$, as the case was for the joint analysis of Fig.~\ref{fig:ALLWI08}. However, the bulk of the data for $T \leq 100$ MeV (represented by the 
shaded bands in Figs.~\ref{fig:PIPELWI08}-\ref{fig:PIMCXWI08}) appears to be either underestimated by the WI08 solution (i.e., in case of the CX reaction) or overestimated by it (i.e., in case of the two ES reactions, the 
effects for the $\pi^+ p$ reaction being more pronounced). One notices that the mismatches decrease with increasing beam energy, converging to $z=1$ in the vicinity of $T=100$ MeV. Such a behavior is consistent with the general 
conclusions of Refs.~\cite{glk,m,mr1} for an energy-dependent isospin-breaking effect.

Of course, the question arises whether the departure from the triangle identity is indicative of missing features in the physical description of the $\pi N$ interaction at low energy. For instance, the assumption has always 
been that the $\pi^+ p$ interaction at low energy is purely elastic and that all the inelasticity corrections, applied to the amplitudes and to the phase shifts, are (small and) known \cite{nord0,nord1,nord2}. On the other 
hand, the speculation is frequently voiced that the mismatches, observed on several occasions in the low-energy $\pi N$ interaction (in particular, in the $\pi^+ p$ reaction), could be attributed to the incompleteness of the 
currently used theoretical background, rather than to experimental discrepancies. For instance, such a scenario could be possible if a (broad) $\pi N$ resonance existed below the $\Delta(1232)$ mass; at present, no such state 
has been detected. Far-fetched as this possibility might appear at first glance, it may be equally difficult to either refute or endorse it.

From Figs.~\ref{fig:PIPELWI08}-\ref{fig:PIMCXWI08} and from Table \ref{tab:WI08Parameters}, one may conclude that the WI08 solution does not describe sufficiently well the bulk of the low-energy measurements in any of the three 
$\pi N$ reactions. Evidently, the WI08 solution at low energy represents a fictitious, average $\pi N$ process, one which does not adequately capture the dynamics of the three physical $\pi N$ reactions. Regarding the analysis 
of the scale factors obtained with the WI08 solution, we stress (once again) that herein we have only fitted straight lines to their $z_j$ values, as they appear in the SAID web page \cite{abws}; we have used no input from our 
approach when analyzing their $z_j$ values.

The conclusion appears to be inevitable. Even in a framework of an analysis assuming the isospin invariance (as the case is for the WI08 solution), the isospin-breaking effects, albeit somewhat hidden, manifest themselves as 
systematic trends in the output of the optimization and may be uncovered if the three $\pi N$ reactions are analyzed \emph{separately}. This conclusion strengthens our argument on investigating the behavior of the normalized 
residuals in PWAs of the low-energy measurements.

One might argue that the effects, contained in Figs.~\ref{fig:PIPELWI08}-\ref{fig:PIMCXWI08} and in Table \ref{tab:WI08Parameters}, are not `large'. However, it must be borne in mind that the isospin-breaking effects in the 
$\pi N$ domain are not expected to be large. Gibbs, Ai, and Kaufmann \cite{glk} have shown that such effects mainly affect the $s$-wave part of the amplitude ($10 \%$ effects are seen in their Fig.~1), and (to a lesser extent) 
the no-spin-flip $p$-wave part (smaller effects are seen in their Figs.~2). We have shown that the largest departure from the isospin invariance occurs in the phase shifts $\delta^{1/2}_{0+}$ and $\delta^{3/2}_{1+}$, see Figs.~2 
and 3 of Ref.~\cite{mr1}. Both analyses appear to agree on the expected magnitude of such effects: roughly speaking, effects between $5$ and $10 \%$ (in the amplitude) are expected at low energy. The systematic effects, observed 
in Figs.~\ref{fig:PIPELWI08}-\ref{fig:PIMCXWI08} of this work, have about the right size, as well as an energy dependence which is consistent with the findings of Refs.~\cite{glk,m,mr1}.

To conclude, we briefly compare the level of the violation of the isospin invariance in the $\pi N$ interaction \cite{glk,m,mr1} with what has long been known for the $N N$ system \cite{mos}. The hadronic part of the low-energy 
$N N$ interaction is characterized by three scattering lengths, corresponding to the three $^1S_0$ states $pp$, $nn$, and $np$. If charge independence (which is used in the $N N$ domain as a synonym for isospin invariance) 
would hold, these three scattering lengths would be equal. In fact, the values obtained after the EM effects are removed are \cite{mos}:
\begin{equation}
a_{pp}=-17.3(4) \,\, {\rm fm}, \, \, \, a_{nn}=-18.8(3) \,\, {\rm fm}, \, \, \, a_{np}=-23.77(9) \,\, {\rm fm} \, \, \, .
\end{equation}
(In Ref.~\cite{mos}, the three scattering lengths carry the superscript `N', indicating that these quantities are nuclear ones, i.e., obtained after the EM corrections have been made.) Obviously, these numbers violate charge 
independence and, to a lesser extent, charge symmetry, as
\begin{equation}
\Delta a_{CD}=(a_{pp}+a_{nn})/2-a_{np}=5.7(3) {\rm fm}
\end{equation}
and
\begin{equation}
\Delta a_{CSD}=a_{pp}-a_{nn}=1.5(5) {\rm fm}
\end{equation}
are significantly non-zero. This corresponds to a violation of the charge independence in the low-energy $N N$ interaction around $27 \%$ and of charge symmetry about $8 \%$. The level of the charge-independence breaking in 
the $N N$ interaction is to be compared with the $5$ to $10 \%$ effects which have been reported in Refs.~\cite{glk,m,mr1} for the low-energy $\pi N$ system. If the $\pi N$ interaction is regarded as the `fundamental' element 
for the description of the $N N$ interaction (as the case is in the framework of the meson-exchange theories of the strong interaction), it is logical to expect that the isospin-breaking effects cascade from the $N N$ 
interaction down to the $\pi N$ interaction. Under the assumption that the nuclear force in the $N N$ interaction is modeled at low energy via the one-pion exchange mechanism, Babenko and Petrov \cite{bp} recently obtained a 
considerable splitting of the $\pi N$ coupling constant, i.e., significantly different values for the couplings of the charged and of the neutral pions to the nucleon. One should not forget that, aiming at providing an 
explanation of the unexpected result of Ref.~\cite{glk}, Piekarewicz had (already in 1995) attributed the isospin-breaking effects to changes in the coupling constant due to the mass difference between the $u$ and the $d$ 
quarks \cite{piek}. In this context, we have also reported changes in the fitted value of the $\pi N$ coupling constant when involving the CX data in the optimization, see Table 4 in Ref.~\cite{mr1}.

The effects, shown in Figs.~\ref{fig:PIPELWI08}-\ref{fig:PIMCXWI08} and in Table \ref{tab:WI08Parameters} of this work, are systematic and indicative of the non-fulfillment of at least one of the three conditions (a)-(c), 
listed in the beginning of this section. One of the safest conclusions, drawn from Figs.~\ref{fig:PIPELKMatrix}-\ref{fig:PIMCXKMatrix}, is that our approach yields results which closely represent and reproduce the bulk of the 
$\pi N$ measurements for $T \leq 100$ MeV; the low-energy behavior of the WI08 solution does not appear to account for the bulk of the low-energy data as successfully. If future investigation in the $\pi N$ sector reveals that 
the meson-factory, low-energy experiments had been affected by severe energy-dependent systematic effects, then a solution based on a dispersion-relation analysis framework (like WI08) may be more reliable than ours. We will 
not speculate on how such effects could affect the two ES processes in one way (i.e., resulting in a systematic underestimation of the relevant DCSs) and the CX reaction in another (i.e., resulting in a systematic overestimation 
of the relevant DCSs). If, on the other hand, the persistent discrepancies, observed in the $\pi N$ interaction at low energy, are to be blamed elsewhere (i.e., on a departure from a theoretical constraint or assumption), then 
our approach will be proven justified.

A final word on the subject of the EM corrections is due. In every PWA of the $\pi N$ data, the goal is the extraction of the hadronic components of the partial-wave amplitudes, and of the hadronic phase shifts derived thereof. 
At the present time, no unique scheme exists for the removal of the EM effects from the $\pi N$ data. Additionally, it is not always clear which EM corrections have been applied in the various PWAs which are available. The 
NORDITA corrections \cite{nord0,nord1,nord2} were obtained in the 1970s, at a time when the meson factories were under construction; as a result, they do not include any of the data sets which became available after the meson 
factories came into operation. Part of our research programme in the late 1990s was the re-assessment of the EM effects at low energy \cite{gmorw1,gmorw2}. To the best of our knowledge, the compatibility of these (and, perhaps, 
also other) EM-correction schemes has not been properly addressed; a comparison has been pursued in Ref.~\cite{ga}, yet only for a small subset of the available low-energy data.

In most works in the $\pi N$ domain, fits of hadronic models are directly performed to the results of the PWAs of the $\pi N$ data (partial-wave amplitudes and phase shifts), assuming that these results represent `purely hadronic' 
quantities. We stress that, so long as residual EM effects are still contained in the important results of the various PWAs (e.g., in the phase-shift solutions), no physical quantities which are estimated on their basis can be 
considered to be purely hadronic. The hope is that the values of the hadronic quantities are not much different from the current estimates, but this hope rests upon the assumption that the residual EM effects in the output of 
the various PWAs are small, which needs checking. As a result, when analyzing the data or when fitting to the output of any PWA of the $\pi N$ data, one must always bear in mind that the available schemes, purporting at the 
removal of the EM effects, are neither complete nor model-independent. We believe that an innovative approach, aiming at the re-assessment of the EM effects in the $\pi N$ domain, would be welcomed. The EM effects must be 
obtained in one consistent scheme for all three reactions and for an extensive energy range, i.e., from the $\pi N$ threshold to the GeV region.

\begin{ack}
We would like to thank I.~I.~Strakovsky for clarifying some questions regarding the output of the SAID Analysis Program. We acknowledge a helpful exchange of electronic mail with G.~A.~Miller on the subject of the violation 
of isospin invariance in the $N N$ interaction.
\end{ack}

\newpage
\begin{table}[h!]
{\bf \caption{\label{tab:Constants}}}The current values of the physical constants, used in this analysis. These values have been taken from the 2014 compilation of the Particle-Data Group \cite{pdg}. Regarding the Roper 
resonance $N(1440)$, the partial width $\Gamma_R$ of Section \ref{sec:KMPIM} is the product of the total width $\Gamma_T$ and the corresponding branching ratio $\eta$ for the $\pi N$ decay modes of the resonance.
\vspace{0.2cm}
\begin{center}
\begin{tabular}{|l|c|}
\hline
Physical quantity (unit) & Value \\
\hline
Charged-pion mass (MeV) & $139.57018$ \\
Proton mass $m_p$ (MeV) & $938.272046$ \\
$\Delta(1232)$ mass $m_{\Delta}$ (MeV) & $1232$ \\
$\Delta(1232)$ decay width $\Gamma_{\Delta}$ (MeV) & $117$ \\
$N(1440)$ $M_R$ (MeV) & $1440$ \\
$N(1440)$ $\Gamma_T$ (MeV) & $300$ \\
$N(1440)$ $\eta$ & $0.650$ \\
\hline
\end{tabular}
\end{center}
\end{table}

\newpage
\begin{table}[h!]
{\bf \caption{\label{tab:DBPIPEL}}}The data sets comprising the truncated $\pi^+ p$ database, the pion laboratory kinetic energy $T_j$ of the data set (in MeV), the number of degrees of freedom $N_j$ after the removal of the 
outliers, the scale factor $z_j$ which minimizes $\chi_j^2$ of Eq.~(\ref{eq:chijsq}), the normalization uncertainty $\delta z_j$ (reported by the experimental group or, if not reported, assigned by us, e.g., as the case is for 
the AULD79 data set), the value of $(\chi_j^2)_{\rm min}$, and the p-value of the fit. In the case of free floating, $z_j$ is equal to $\hat{z}_j$ of Eq.~(\ref{eq:zjopt}). The numbers in this table correspond to the final fit 
to the data using the $K$-matrix parameterizations of Section \ref{sec:KMPIPEL}.
\vspace{0.2cm}
\begin{center}
\begin{tabular}{|l|c|c|c|c|c|c|l|}
\hline
Data set & $T_j$ & $N_j$ & $z_j$ & $\delta z_j$ & $(\chi_j^2)_{\rm min}$ & p-value & Comments \\
\hline
\multicolumn{8}{|c|}{Differential cross sections} \\
\hline
AULD79 & $47.90$ & $11$ & $1.0290$ & $0.1097$ & $16.4530$ & $0.1251$ & \\
RITCHIE83 & $65.00$ & $8$ & $1.0469$ & $0.0240$ & $18.2215$ & $0.0196$ & \\
RITCHIE83 & $72.50$ & $10$ & $1.0062$ & $0.0200$ & $4.7931$ & $0.9046$ & \\
RITCHIE83 & $80.00$ & $10$ & $1.0289$ & $0.0140$ & $18.9749$ & $0.0406$ & \\
RITCHIE83 & $95.00$ & $10$ & $1.0308$ & $0.0150$ & $12.1585$ & $0.2746$ & \\
FRANK83 & $29.40$ & $28$ & $1.0423$ & $0.0370$ & $19.3186$ & $0.8880$ & \\
FRANK83 & $49.50$ & $28$ & $1.0581$ & $0.2030$ & $34.4110$ & $0.1877$ & \\
FRANK83 & $69.60$ & $27$ & $0.9304$ & $0.0950$ & $23.2937$ & $0.6691$ & \\
FRANK83 & $89.60$ & $27$ & $0.8603$ & $0.0470$ & $29.1001$ & $0.3561$ & \\
BRACK86 & $66.80$ & $4$ & $0.8941$ & $0.0120$ & $2.4056$ & $0.6616$ & freely floated \\
BRACK86 & $86.80$ & $8$ & $0.9368$ & $0.0140$ & $16.3766$ & $0.0373$ & freely floated \\
BRACK86 & $91.70$ & $5$ & $0.9726$ & $0.0120$ & $12.5697$ & $0.0278$ & \\
BRACK86 & $97.90$ & $5$ & $0.9709$ & $0.0150$ & $7.6323$ & $0.1777$ & \\
BRACK88 & $66.80$ & $6$ & $0.9488$ & $0.0210$ & $10.6388$ & $0.1002$ & \\
BRACK88 & $66.80$ & $6$ & $0.9574$ & $0.0210$ & $9.3760$ & $0.1535$ & \\
WIEDNER89 & $54.30$ & $19$ & $0.9894$ & $0.0304$ & $14.8601$ & $0.7314$ & \\
BRACK90 & $30.00$ & $5$ & $1.1805$ & $0.0360$ & $8.4943$ & $0.1310$ & freely floated \\
BRACK90 & $45.00$ & $8$ & $1.0258$ & $0.0220$ & $8.4449$ & $0.3913$ & \\
BRACK95 & $87.10$ & $8$ & $0.9717$ & $0.0220$ & $13.8636$ & $0.0854$ & \\
BRACK95 & $98.10$ & $8$ & $0.9796$ & $0.0200$ & $15.0499$ & $0.0582$ & \\
JORAM95 & $45.10$ & $9$ & $0.9712$ & $0.0330$ & $17.8659$ & $0.0368$ & $124.42^\circ$ removed \\
JORAM95 & $68.60$ & $9$ & $1.0516$ & $0.0440$ & $9.1437$ & $0.4241$ & \\
JORAM95 & $32.20$ & $20$ & $1.0224$ & $0.0340$ & $31.2306$ & $0.0522$ & \\
JORAM95 & $44.60$ & $18$ & $0.9621$ & $0.0340$ & $27.6309$ & $0.0679$ & $30.74^\circ$, $35.40^\circ$ removed \\
\hline
\end{tabular}
\end{center}
\end{table}

\newpage
\begin{table*}
{\bf Table 2 continued}
\vspace{0.2cm}
\begin{center}
\begin{tabular}{|l|c|c|c|c|c|c|l|}
\hline
Data set & $T_j$ & $N_j$ & $z_j$ & $\delta z_j$ & $(\chi_j^2)_{\rm min}$ & p-value & Comments \\
\hline
\multicolumn{8}{|c|}{Analyzing powers} \\
\hline
SEVIOR89 & $98.00$ & $6$ & $1.0128$ & $0.0740$ & $5.5312$ & $0.4777$ & \\
WIESER96 & $68.34$ & $3$ & $0.9100$ & $0.0500$ & $4.5009$ & $0.2122$ & \\
WIESER96 & $68.34$ & $4$ & $0.9316$ & $0.0500$ & $4.7278$ & $0.3164$ & \\
MEIER04 & $57.20$-$87.20$ & $12$ & $0.9824$ & $0.0350$ & $13.9222$ & $0.3057$ & \\
MEIER04 & $45.20$, $51.20$ & $6$ & $0.9709$ & $0.0350$ & $7.4638$ & $0.2801$ & \\
MEIER04 & $57.30$-$87.20$ & $7$ & $1.0038$ & $0.0350$ & $11.0607$ & $0.1360$ & \\
\hline
\multicolumn{8}{|c|}{Partial-total cross sections} \\
\hline
KRISS97 & $39.80$ & $1$ & $1.0144$ & $0.0300$ & $2.5847$ & $0.1079$ & \\
KRISS97 & $40.50$ & $1$ & $1.0023$ & $0.0300$ & $0.2646$ & $0.6070$ & \\
KRISS97 & $44.70$ & $1$ & $1.0042$ & $0.0300$ & $0.1939$ & $0.6597$ & \\
KRISS97 & $45.30$ & $1$ & $1.0053$ & $0.0300$ & $0.2402$ & $0.6241$ & \\
KRISS97 & $51.10$ & $1$ & $1.0265$ & $0.0300$ & $4.1742$ & $0.0410$ & \\
KRISS97 & $51.70$ & $1$ & $1.0046$ & $0.0300$ & $0.1507$ & $0.6979$ & \\
KRISS97 & $54.80$ & $1$ & $1.0111$ & $0.0300$ & $0.3650$ & $0.5457$ & \\
KRISS97 & $59.30$ & $1$ & $1.0294$ & $0.0300$ & $1.7134$ & $0.1905$ & \\
KRISS97 & $66.30$ & $2$ & $1.0523$ & $0.0300$ & $4.4518$ & $0.1080$ & \\
KRISS97 & $66.80$ & $2$ & $1.0077$ & $0.0300$ & $0.6370$ & $0.7272$ & \\
KRISS97 & $80.00$ & $1$ & $1.0135$ & $0.0300$ & $0.3320$ & $0.5645$ & \\
KRISS97 & $89.30$ & $1$ & $1.0074$ & $0.0300$ & $0.2509$ & $0.6164$ & \\
KRISS97 & $99.20$ & $1$ & $1.0541$ & $0.0300$ & $3.9747$ & $0.0462$ & \\
FRIEDMAN99 & $45.00$ & $1$ & $1.0467$ & $0.0600$ & $2.7169$ & $0.0993$ & \\
FRIEDMAN99 & $52.10$ & $1$ & $1.0217$ & $0.0600$ & $0.3985$ & $0.5279$ & \\
FRIEDMAN99 & $63.10$ & $1$ & $1.0401$ & $0.0600$ & $0.5992$ & $0.4389$ & \\
FRIEDMAN99 & $67.45$ & $2$ & $1.0546$ & $0.0600$ & $1.3394$ & $0.5119$ & \\
FRIEDMAN99 & $71.50$ & $2$ & $1.0508$ & $0.0600$ & $0.8682$ & $0.6479$ & \\
FRIEDMAN99 & $92.50$ & $2$ & $1.0411$ & $0.0600$ & $0.5389$ & $0.7638$ & \\
\hline
\end{tabular}
\end{center}
\end{table*}

\newpage
\begin{table*}
{\bf Table 2 continued}
\vspace{0.2cm}
\begin{center}
\begin{tabular}{|l|c|c|c|c|c|c|l|}
\hline
Data set & $T_j$ & $N_j$ & $z_j$ & $\delta z_j$ & $(\chi_j^2)_{\rm min}$ & p-value & Comments \\
\hline
\multicolumn{8}{|c|}{Total-nuclear cross sections} \\
\hline
CARTER71 & $71.60$ & $1$ & $1.0940$ & $0.0600$ & $2.7880$ & $0.0950$ & \\
CARTER71 & $97.40$ & $1$ & $1.0475$ & $0.0600$ & $0.6326$ & $0.4264$ & \\
PEDRONI78 & $72.50$ & $1$ & $1.0126$ & $0.0600$ & $0.1451$ & $0.7033$ & \\
PEDRONI78 & $84.80$ & $1$ & $1.0302$ & $0.0600$ & $0.3085$ & $0.5786$ & \\
PEDRONI78 & $95.10$ & $1$ & $1.0215$ & $0.0600$ & $0.1766$ & $0.6743$ & \\
PEDRONI78 & $96.90$ & $1$ & $1.0154$ & $0.0600$ & $0.1132$ & $0.7365$ & \\
\hline
\end{tabular}
\end{center}
\end{table*}

\newpage
\begin{table}[h!]
{\bf \caption{\label{tab:DBPIMEL}}}The equivalent of Table \ref{tab:DBPIPEL} for the fits to our $\pi^- p$ elastic-scattering database. The numbers of this table correspond to the final fit to the data using the $K$-matrix 
parameterizations of Section \ref{sec:KMPIM}.
\vspace{0.2cm}
\begin{center}
\begin{tabular}{|l|c|c|c|c|c|c|l|}
\hline
Data set & $T_j$ & $N_j$ & $z_j$ & $\delta z_j$ & $(\chi_j^2)_{\rm min}$ & p-value & Comments \\
\hline
\multicolumn{8}{|c|}{Differential cross sections} \\
\hline
FRANK83 & $29.40$ & $28$ & $0.9841$ & $0.0350$ & $30.9946$ & $0.3173$ & \\
FRANK83 & $49.50$ & $28$ & $1.0985$ & $0.0780$ & $29.4176$ & $0.3916$ & \\
FRANK83 & $69.60$ & $27$ & $1.0921$ & $0.2530$ & $24.7552$ & $0.5882$ & \\
FRANK83 & $89.60$ & $27$ & $0.9457$ & $0.1390$ & $24.8611$ & $0.5822$ & \\
BRACK86 & $66.80$ & $5$ & $0.9974$ & $0.0130$ & $13.9614$ & $0.0159$ & \\
BRACK86 & $86.80$ & $5$ & $1.0031$ & $0.0120$ & $1.3828$ & $0.9262$ & \\
BRACK86 & $91.70$ & $5$ & $0.9962$ & $0.0120$ & $3.0724$ & $0.6888$ & \\
BRACK86 & $97.90$ & $5$ & $0.9998$ & $0.0120$ & $5.9607$ & $0.3101$ & \\
WIEDNER89 & $54.30$ & $18$ & $1.1567$ & $0.0304$ & $23.5424$ & $0.1706$ & $15.55^\circ$ removed, freely floated \\
BRACK90 & $30.00$ & $5$ & $1.0210$ & $0.0200$ & $5.0513$ & $0.4096$ & \\
BRACK90 & $45.00$ & $9$ & $1.0512$ & $0.0220$ & $11.6945$ & $0.2311$ & \\
BRACK95 & $87.50$ & $6$ & $0.9812$ & $0.0220$ & $10.6655$ & $0.0993$ & \\
BRACK95 & $98.10$ & $7$ & $1.0072$ & $0.0210$ & $8.4645$ & $0.2934$ & $36.70^\circ$ removed \\
JORAM95 & $32.70$ & $4$ & $0.9941$ & $0.0330$ & $3.9286$ & $0.4158$ & \\
JORAM95 & $32.70$ & $2$ & $0.9533$ & $0.0330$ & $5.6499$ & $0.0593$ & \\
JORAM95 & $45.10$ & $4$ & $0.9540$ & $0.0330$ & $12.5594$ & $0.0136$ & \\
JORAM95 & $45.10$ & $3$ & $0.9450$ & $0.0330$ & $9.5406$ & $0.0229$ & \\
JORAM95 & $68.60$ & $7$ & $1.0829$ & $0.0440$ & $14.3284$ & $0.0456$ & \\
JORAM95 & $68.60$ & $3$ & $1.0322$ & $0.0440$ & $2.3004$ & $0.5124$ & \\
JORAM95 & $32.20$ & $20$ & $1.0615$ & $0.0340$ & $21.3464$ & $0.3770$ & \\
JORAM95 & $44.60$ & $20$ & $0.9434$ & $0.0340$ & $30.3198$ & $0.0648$ & \\
JANOUSCH97 & $43.60$ & $1$ & $1.0395$ & $0.1500$ & $0.1482$ & $0.7002$ & \\
JANOUSCH97 & $50.30$ & $1$ & $1.0320$ & $0.1500$ & $0.1189$ & $0.7302$ & \\
JANOUSCH97 & $57.30$ & $1$ & $1.0829$ & $0.1500$ & $4.7450$ & $0.0294$ & \\
JANOUSCH97 & $64.50$ & $1$ & $1.0212$ & $0.1500$ & $0.0298$ & $0.8629$ & \\
JANOUSCH97 & $72.00$ & $1$ & $1.2988$ & $0.1500$ & $4.6476$ & $0.0311$ & \\
\hline
\end{tabular}
\end{center}
\end{table}

\newpage
\begin{table*}
{\bf Table 3 continued}
\vspace{0.2cm}
\begin{center}
\begin{tabular}{|l|c|c|c|c|c|c|l|}
\hline
Data set & $T_j$ & $N_j$ & $z_j$ & $\delta z_j$ & $(\chi_j^2)_{\rm min}$ & p-value & Comments \\
\hline
\multicolumn{8}{|c|}{Analyzing powers} \\
\hline
ALDER83 & $98.00$ & $6$ & $1.0103$ & $0.0400$ & $5.3078$ & $0.5050$ & \\
SEVIOR89 & $98.00$ & $5$ & $0.9871$ & $0.0740$ & $1.7012$ & $0.8887$ & \\
HOFMAN98 & $86.80$ & $11$ & $1.0019$ & $0.0300$ & $5.9941$ & $0.8738$ & \\
PATTERSON02 & $57.20$ & $10$ & $0.9468$ & $0.0370$ & $10.5993$ & $0.3896$ & \\
PATTERSON02 & $66.90$ & $9$ & $0.9988$ & $0.0370$ & $4.6096$ & $0.8669$ & \\
PATTERSON02 & $66.90$ & $10$ & $0.9522$ & $0.0370$ & $16.4217$ & $0.0882$ & \\
PATTERSON02 & $87.20$ & $11$ & $0.9835$ & $0.0370$ & $8.1202$ & $0.7025$ & \\
PATTERSON02 & $87.20$ & $11$ & $0.9944$ & $0.0370$ & $4.9386$ & $0.9341$ & \\
PATTERSON02 & $98.00$ & $12$ & $0.9950$ & $0.0370$ & $6.5960$ & $0.8831$ & \\
MEIER04 & $67.30$, $87.20$ & $3$ & $0.9929$ & $0.0350$ & $3.2264$ & $0.3580$ & \\
\hline
\end{tabular}
\end{center}
\end{table*}

\newpage
\begin{table}[h!]
{\bf \caption{\label{tab:DBPIMCX}}}The equivalent of Table \ref{tab:DBPIPEL} for the fits to our $\pi^- p$ charge-exchange database. The numbers of this table correspond to the final fit to the data using the $K$-matrix 
parameterizations of Section \ref{sec:KMPIM}.
\vspace{0.2cm}
\begin{center}
\begin{tabular}{|l|c|c|c|c|c|c|l|}
\hline
Data set & $T_j$ & $N_j$ & $z_j$ & $\delta z_j$ & $(\chi_j^2)_{\rm min}$ & p-value & Comments \\
\hline
\multicolumn{8}{|c|}{Differential cross sections} \\
\hline
DUCLOS73 & $22.60$ & $1$ & $0.9422$ & $0.0800$ & $1.2111$ & $0.2711$ & \\
DUCLOS73 & $32.90$ & $1$ & $0.9715$ & $0.0800$ & $0.2741$ & $0.6006$ & \\
DUCLOS73 & $42.60$ & $1$ & $0.9100$ & $0.0800$ & $2.3763$ & $0.1232$ & \\
FITZGERALD86 & $32.48$ & $2$ & $1.5008$ & $0.0780$ & $2.3360$ & $0.3110$ & freely floated \\
FITZGERALD86 & $36.11$ & $2$ & $1.7161$ & $0.0780$ & $1.2144$ & $0.5449$ & freely floated \\
FITZGERALD86 & $40.26$ & $2$ & $1.8332$ & $0.0780$ & $6.5440$ & $0.0379$ & freely floated \\
FITZGERALD86 & $47.93$ & $2$ & $1.4509$ & $0.0780$ & $1.6502$ & $0.4382$ & freely floated \\
FITZGERALD86 & $51.78$ & $3$ & $1.1216$ & $0.0780$ & $7.2398$ & $0.0646$ & \\
FITZGERALD86 & $55.58$ & $3$ & $1.0933$ & $0.0780$ & $2.5533$ & $0.4657$ & \\
FITZGERALD86 & $63.21$ & $3$ & $1.0548$ & $0.0780$ & $1.3372$ & $0.7203$ & \\
FRLEZ98 & $27.50$ & $6$ & $1.0904$ & $0.0870$ & $10.4326$ & $0.1076$ & \\
ISENHOWER99 & $10.60$ & $4$ & $1.0196$ & $0.0600$ & $2.1628$ & $0.7058$ & \\
ISENHOWER99 & $10.60$ & $5$ & $1.0054$ & $0.0400$ & $1.4569$ & $0.9180$ & \\
ISENHOWER99 & $10.60$ & $6$ & $1.0176$ & $0.0400$ & $8.0670$ & $0.2332$ & \\
ISENHOWER99 & $20.60$ & $5$ & $0.9799$ & $0.0400$ & $1.5691$ & $0.9050$ & \\
ISENHOWER99 & $20.60$ & $6$ & $1.0120$ & $0.0400$ & $8.1842$ & $0.2249$ & \\
ISENHOWER99 & $39.40$ & $4$ & $1.0707$ & $0.0600$ & $7.3129$ & $0.1202$ & \\
ISENHOWER99 & $39.40$ & $5$ & $1.0593$ & $0.0400$ & $8.4695$ & $0.1322$ & \\
ISENHOWER99 & $39.40$ & $5$ & $0.9521$ & $0.0400$ & $5.1290$ & $0.4003$ & \\
SADLER04 & $63.86$ & $20$ & $0.9540$ & $0.0650$ & $16.2394$ & $0.7017$ & \\
SADLER04 & $83.49$ & $20$ & $0.9877$ & $0.0520$ & $11.7013$ & $0.9260$ & \\
SADLER04 & $94.57$ & $20$ & $1.0303$ & $0.0450$ & $7.2943$ & $0.9956$ & \\
JIA08 & $34.37$ & $4$ & $0.8433$ & $0.1000$ & $4.9400$ & $0.2935$ & \\
JIA08 & $39.95$ & $4$ & $0.8676$ & $0.1000$ & $3.1953$ & $0.5257$ & \\
JIA08 & $43.39$ & $4$ & $0.8780$ & $0.1000$ & $2.5106$ & $0.6427$ & \\
JIA08 & $46.99$ & $4$ & $0.9799$ & $0.1000$ & $5.1602$ & $0.2713$ & \\
\hline
\end{tabular}
\end{center}
\end{table}

\newpage
\begin{table*}
{\bf Table 4 continued}
\vspace{0.2cm}
\begin{center}
\begin{tabular}{|l|c|c|c|c|c|c|l|}
\hline
Data set & $T_j$ & $N_j$ & $z_j$ & $\delta z_j$ & $(\chi_j^2)_{\rm min}$ & p-value & Comments \\
\hline
\multicolumn{8}{|c|}{Differential cross sections} \\
\hline
JIA08 & $54.19$ & $4$ & $0.9057$ & $0.1000$ & $2.1191$ & $0.7139$ & \\
JIA08 & $59.68$ & $4$ & $0.9318$ & $0.1000$ & $3.0789$ & $0.5447$ & \\
MEKTEROVIC09 & $33.89$ & $20$ & $1.0239$ & $0.0340$ & $17.0108$ & $0.6523$ & \\
MEKTEROVIC09 & $39.38$ & $20$ & $1.0145$ & $0.0260$ & $14.7679$ & $0.7895$ & \\
MEKTEROVIC09 & $44.49$ & $20$ & $1.0100$ & $0.0270$ & $33.1582$ & $0.0324$ & \\
MEKTEROVIC09 & $51.16$ & $20$ & $1.0355$ & $0.0290$ & $15.0640$ & $0.7727$ & \\
MEKTEROVIC09 & $57.41$ & $20$ & $1.0390$ & $0.0290$ & $19.4998$ & $0.4896$ & \\
MEKTEROVIC09 & $66.79$ & $20$ & $1.0227$ & $0.0300$ & $19.5380$ & $0.4871$ & \\
MEKTEROVIC09 & $86.62$ & $20$ & $1.0016$ & $0.0290$ & $31.0086$ & $0.0551$ & \\
\hline
\multicolumn{8}{|c|}{Legendre expansion of the differential cross section} \\
\hline
SALOMON84 & $27.40$ & $3$ & $0.9722$ & $0.0310$ & $2.7669$ & $0.4290$ & \\
SALOMON84 & $39.30$ & $3$ & $0.9939$ & $0.0310$ & $1.0926$ & $0.7789$ & \\
BAGHERI88 & $45.60$ & $3$ & $1.0055$ & $0.0310$ & $0.1372$ & $0.9870$ & \\
BAGHERI88 & $62.20$ & $3$ & $0.9582$ & $0.0310$ & $3.6125$ & $0.3065$ & \\
BAGHERI88 & $76.40$ & $3$ & $0.9725$ & $0.0310$ & $3.3190$ & $0.3450$ & \\
BAGHERI88 & $91.70$ & $3$ & $1.0151$ & $0.0310$ & $2.8312$ & $0.4184$ & \\
\hline
\multicolumn{8}{|c|}{Measurement of the width of pionic hydrogen} \\
\hline
SCHROEDER01 & $0.00$ & $1$ & $0.9746$ & $0.0225$ & $2.5283$ & $0.1118$ & \\
\hline
\multicolumn{8}{|c|}{Analyzing powers} \\
\hline
STASKO93 & $100.00$ & $4$ & $0.9944$ & $0.0440$ & $1.4508$ & $0.8353$ & \\
GAULARD99 & $98.10$ & $6$ & $1.0234$ & $0.0450$ & $1.0835$ & $0.9823$ & \\
\hline
\end{tabular}
\end{center}
\end{table*}

\newpage
\begin{table*}
{\bf Table 4 continued}
\vspace{0.2cm}
\begin{center}
\begin{tabular}{|l|c|c|c|c|c|c|l|}
\hline
Data set & $T_j$ & $N_j$ & $z_j$ & $\delta z_j$ & $(\chi_j^2)_{\rm min}$ & p-value & Comments \\
\hline
\multicolumn{8}{|c|}{Total cross sections} \\
\hline
BUGG71 & $90.90$ & $1$ & $1.0226$ & $0.0600$ & $0.1478$ & $0.7007$ & \\
BREITSCHOPF06 & $38.90$ & $1$ & $0.9960$ & $0.0300$ & $0.1641$ & $0.6854$ & \\
BREITSCHOPF06 & $43.00$ & $1$ & $1.0011$ & $0.0300$ & $0.0259$ & $0.8722$ & \\
BREITSCHOPF06 & $47.10$ & $1$ & $0.9980$ & $0.0300$ & $0.0578$ & $0.8100$ & \\
BREITSCHOPF06 & $55.60$ & $1$ & $0.9951$ & $0.0300$ & $0.2131$ & $0.6444$ & \\
BREITSCHOPF06 & $64.30$ & $1$ & $0.9724$ & $0.0300$ & $3.8191$ & $0.0507$ & \\
BREITSCHOPF06 & $65.90$ & $1$ & $0.9777$ & $0.0300$ & $2.3820$ & $0.1227$ & \\
BREITSCHOPF06 & $76.10$ & $1$ & $0.9812$ & $0.0300$ & $1.6410$ & $0.2002$ & \\
BREITSCHOPF06 & $96.50$ & $1$ & $0.9819$ & $0.0300$ & $0.5924$ & $0.4415$ & \\
\hline
\end{tabular}
\end{center}
\end{table*}

\newpage
\begin{table}[h!]
{\bf \caption{\label{tab:WI08Parameters}}}The fitted values of the parameters of the weighted linear fit to the data shown in Figs.~\ref{fig:PIPELWI08}-\ref{fig:ALLWI08}, as well as their uncertainties, corrected with the Birge 
factor $\sqrt{\chi^2/{\rm NDF}}$, taking account of the goodness of each fit.
\vspace{0.2cm}
\begin{center}
\begin{tabular}{|l|c|c|}
\hline
Reaction & Intercept & Slope ($10^{-3}$ MeV$^{-1}$) \\
\hline
$\pi^+ p$ & $0.936 \pm 0.028$ & $0.70 \pm 0.34$ \\
$\pi^- p$ elastic-scattering & $0.972 \pm 0.021$ & $0.16 \pm 0.27$ \\
$\pi^- p$ charge-exchange & $1.041 \pm 0.019$ & $-0.43 \pm 0.34$ \\
All three $\pi N$ reactions & $0.985 \pm 0.014$ & $0.10 \pm 0.18$ \\
\hline
\end{tabular}
\end{center}
\end{table}

\clearpage

\clearpage

\clearpage

\clearpage

\clearpage

\clearpage

\clearpage

\end{document}